\documentstyle[preprint,prl,aps]{revtex}
\begin{document}
\draft
\title{Quantum Statistical Mechanics for Nonextensive Systems II}
\author{E. K. Lenzi$^{1}$,  R. S. Mendes$^{2}$ and A. K. Rajagopal$^{3}$ }
\address{$^1$Centro Brasileiro de Pesquisas F\'\i sicas, R. Dr.  Xavier Sigaud 150,
\\22290-180 Rio de Janeiro-RJ, Brazil\\
$^2$Departamento de F\'\i sica, Universidade Estadual de
Maring\'a, \\ Av. Colombo 5790, 87020-900 Maring\'a-PR, Brazil
\\$^3$Naval Research Laboratory, Washington DC 20375-5320 \\}
\date{\today }
\maketitle
\begin{abstract}

In this paper, the Green function theory of quantum many-particle
systems recently presented is reworked within the framework of
nonextensive statistical mechanics with a new normalized
$q$-expectation values. This reformulation introduces a
renormalization of temperature of the earlier theory and a
self-consistency condition. The linear response theory is also
presented, along with its two-particle Green function version.
Finally, a Boltzmann transport-like equation is also developed
here.

\end{abstract}
\pacs{PACS number(s): 05.70.Ce, 05.30.-d, 05.20.-y, 05.30.Ch}
\date{\today}

{\section {INTRODUCTION} }

In two recent papers \cite{RAJA,RAJA2} a Green function theory of
nonextensive many-particle systems was developed. This work is
here reformulated in terms of a new version \cite{NORMA} of the
Tsallis ensemble involving {\it normalized} $q$-expectation values
in the generalized context. This new version obtained in this way
has the following desirable properties which the early versions
lacked\cite{T88,CT91}: invariance of the ensemble with respect to
uniform translation of the energy spectrum, the expectation value
of a $c$-number is the same as that $c$-number and finally, the
preservation of the formal additive structure of an operator
associated with two subsystems (e.g., energy). All these features
entail the appearance of renormalized temperature  parameter
replacing that in the previous formalism along with a
self-consistent condition. The purpose this paper is to revisit
the general Green function theory presented in \cite{RAJA,RAJA2}
in terms of the new framework. In Sec. II the one-particle Green
function theory for both the Bose and Fermi systems along with
explicit results for free-particle systems is given. These have
implications to experiments as described before \cite{RAJA,RAJA2}.
In Sec. III a generalization of the linear response theory
\cite{R96} in the new framework is given and it is re-expressed in
some special cases in terms of the two-particle Green-function.
This is related to scattering cross section as before
\cite{RAJA,RAJA2}. In Sec. IV a Boltzmann transport-like equation
is, for the first time, derived, for arbitray $q$. The final Sec.
V contains conclusions of this work.

${ }$ {\section {THE GREEN FUNCTION THEORY} } ${ }$

The Tsallis entropy $ S_q = (1-{\mbox {Tr}}\hat{\rho}^q)/(q-1)$,
${\mbox {Tr}} \hat{\rho} =1$, $\hat{\rho}$ the system density
matrix is maximized subject to normalized $q$-expectations values
$\langle \hat{H} \rangle_{q}$$ =$$ \tilde{U}_q $ $=$ ${\mbox {Tr}}
\hat{H} \hat{\rho}^q/ {\mbox {Tr}} \hat{\rho}^q  $ , $ \langle
\hat{N} \rangle_{q}$$ =$$ \tilde{N}_q$$=$${\mbox {Tr}} \hat{N}
\hat{\rho}^q / {\mbox {Tr}} \hat{\rho}^q $. A detailed discussion
of this in relation to the earlier formulations may be found in
\cite{NORMA}. To distinguish from our earlier work, a tilda over
the corresponding expressions in the foregoing sequel is used. The
new density matrix determined in this way is then
\begin{eqnarray}
\hat{\tilde{\rho}} &=& \exp_q \left( -\hat{X}(\tilde{\beta},\mu)
\right)/ \tilde{Z}_q \;\;, \nonumber \\ \tilde{Z}_q &=& {\mbox
{Tr}} \left[ \exp_q \left( -\hat{X}(\tilde{\beta},\mu) \right)
\right]\;\;, \label{1qq}
\end{eqnarray}
where
\begin{eqnarray}
\exp_q \left( -\hat{X}(\tilde{\beta},\mu) \right) &\equiv&
\left(1-(1-q) \hat{X}(\tilde{\beta},\mu) \right)^{1/(1-q)} \;,
\nonumber \\ X \left( \tilde{\beta}, \mu \right)&=&
\tilde{\beta}\left( \left(\hat{H}-\tilde{U}_q \right)-\mu
\left(\hat{N}- \tilde{N}_q \right) \right) \;,
\end{eqnarray}
with $\tilde{\beta}=\beta / c_q$, and $c_q={\mbox
{Tr}}\hat{\tilde{\rho}}^{\;q}$. Here $\tilde{\beta}$ is the
normalized temperature. $\beta$ and $\beta \mu$ are the Lagrange
multipliers as before associated with the normalized
$q$-expectation values given above. The constant $c_q$ is a
function of all these quantities.  There are two self-consistency
conditions that follow from Eq.(\ref{1qq}), because
\begin{eqnarray}
\left( \tilde{Z}_q \hat{\tilde{\rho}}\right)^{1-q} &=&\left [
1-(1-q)\tilde{\beta}((\hat{H}- \tilde{U}_q) - \mu (\hat{N} -
\tilde{N}_q )) \right ], \label{2qq}
\end{eqnarray}
from which it follows that
\begin{eqnarray}
c_q=\tilde{Z}_q^{1-q} \;. \label{2.2qq}
\end{eqnarray}
By using the expression for $\hat{\tilde{\rho}}$ in Eq.(\ref{1qq})
in the definition of $c_q$, another relation is derived:
\begin{eqnarray}
c_q=\left.{\mbox {Tr}}\left[ \exp_q \left(
-\hat{X}(\tilde{\beta},\mu)\right)\right]^q \right/\tilde{Z}_q^q
\;\;. \label{3qq}
\end{eqnarray}
Using Eq.(\ref{2.2qq}) in Eq.(\ref{3qq}) an equivalent expression
is obtained for $\tilde{Z}_q$:
\begin{eqnarray}\tilde{Z}_q ={\mbox Tr}
\left[ \exp_q \left(
-\hat{X}(\tilde{\beta},\mu)\right)\right]^q\;. \label{4qq}
\end{eqnarray}
Noting that $\exp_q\left(-\hat{X}(\tilde{\beta},\mu)\right) \equiv
\left[\exp_q\left(-\hat{X}(\tilde{\beta},\mu) \right) \right]^q
\left \{ 1-(1-q)\hat{X}(\tilde{\beta}, \mu)\right \}$ and since
${\mbox {Tr}} \left[\exp_q\left(-\hat{X}(\tilde{\beta},\mu)
\right) \right]^q\hat{X} \equiv 0$ by definition, these two
expression for $\tilde{Z}_q$ become a tautology. The expression
for a general normalized $q$-expectation value is then
\begin{eqnarray}
\langle \hat{A} \rangle_q &=& \tilde{A}_q = \frac{ {\mbox
{Tr}}\hat{A} \hat{\tilde{\rho}}^q} {{\mbox
{Tr}}\hat{\tilde{\rho}}^q}\nonumber \\ &=&\left.{\mbox
{Tr}}\hat{A} \left[ \exp_q \left(
-\hat{X}(\tilde{\beta},\mu)\right)\right]^q \right /c_q
\tilde{Z}_q^{q} \nonumber \\ &=& \frac{ {\mbox {Tr}} \left \{
\hat{A} \left[ \exp_q \left(
-\hat{X}(\tilde{\beta},\mu)\right)\right]^q \right \} }{ {\mbox
{Tr}} \left[ \exp_q \left(
-\hat{X}(\tilde{\beta},\mu)\right)\right]^q } \;\;. \label{5qq}
\end{eqnarray}
Here Eqs.(\ref{2.2qq},\ref{3qq}) were used. From Eq.(\ref{5qq}),
$\tilde{U}_q$, $\tilde{N}_q$ are deduced which along with
Eqs.(\ref{1qq},\ref{2.2qq},\ref{3qq},\ref{4qq}) form the required
self-consistent relations. In this form for the expectation value,
only the ``connected'' diagrams appear and thus explain the
results found \cite{Abe1,Abe2} contract in to that obtained in
\cite{A1} which did not used normalized $q$-expectation values.
The one-particle Green function is then redefined as in
\cite{RAJA2} with the new normalized $q$-average and the density
matrix given above:
\begin{eqnarray}
\tilde{G}^{(q)}(1, 1'; \beta, \mu) &=& \frac{1}{i} \langle {\mbox
{T}} ( \Psi(1) \Psi^{\dagger}(1') )\rangle_q \;,
 \nonumber \\&\equiv & \frac{1}{i }
\frac{{\mbox {Tr}} \left[\hat{\tilde{\rho}}^q {\mbox {T}}( \Psi(1)
\Psi^{\dagger}(1') )\right]} {{\mbox {Tr}}\hat{\tilde{\rho}}^q}
\;\;. \label{6qq}
\end{eqnarray}
Using the contour integral representation employed in
\cite{RAJA,RAJA2}
\begin{eqnarray}
 b^{1-z} \frac{i}{2 \pi }
\int_C {\mbox {du}}\;\exp(- u\;b)(-u)^{-z}= \frac{1}{\Gamma(z)}
\label{7qq}
\end{eqnarray}
with $ b>0$ and ${\mbox {Re}}\; z>0$, where the contour $C$ starts
from $ +\infty$  on the real axis, encircles the origin once
counterclockwise and returns to $ +\infty$, Eq.(\ref{6qq}) is
expressed in terms of the Green function with $q=1$:
\begin{eqnarray}
\tilde{G}^{(q)}(1,1' ;\beta, \mu )=\int_{C} {\mbox {d}}u
\tilde{K}_q^{(2)}(u) \tilde{Z}_{1}(-u(1-q)\tilde{\beta}, \mu )\;
G^{(1)}(1,1';-u(1-q)\tilde{\beta}, \mu) \;\;, \label{8.1qq}
\end{eqnarray}
where
\begin{eqnarray}\tilde{K}_q^{(2)}(u)&=& i
\frac{\Gamma(1/(1-q))}{2\pi\; \tilde{Z}_q} \exp(-u)(-u)^{1/(1-q)}
\nonumber \\ 1&=& \int_C {\mbox {d}}u \tilde{K}^{(2)}_q(u)
\tilde{Z}_{1}(-u(1-q)\tilde{\beta}, \mu ) \;\;. \label{8.2qq}
\end{eqnarray}
Note that all these definitions are consistent with Eq.(\ref{1qq})
defined above. Note also that these expressions have similar
appearance as those found in \cite{RAJA2}. When $q=1$, the
constant terms involving, $\tilde{U}_q$, $\tilde{N}_q$ factor out
so that the one-particle Green function in integrand of
Eq.(\ref{8.1qq}) is the conventional  extensive one: $\tilde{Z}_1$
appearing there however is the corresponding conventional
partition function multiplied by $e^{-u\tilde{\beta}(1-q)
(\tilde{U}_q -\mu \tilde{N}_q)}$. From Eq.(\ref{1qq}) an
equivalent expression is found for $\tilde{Z}_q$:
\begin{eqnarray}
\tilde{Z}_q &=& \int_C {\mbox {d}}u K_q^{(1)}(u)\tilde{Z}_1(-u
(1-q) \tilde{\beta}, \mu) \nonumber \\
K_q^{(1)}&=&\frac{i\Gamma\left(\frac{2-q}{1-q} \right)}{2 \pi}
\exp(-u)(-u)^{-(2-q)/(1-q)} \nonumber \\ &=&
\frac{\tilde{Z}_q}{u(1-q)} \tilde{K}_q^{(2)}(u) \;\;.
\label{8.3qq}
\end{eqnarray}
As in \cite{RAJA2}, correlation functions are introduced.
Following KB \cite{K62}, we introduce correlation functions
\begin{eqnarray}
\tilde{G}_{>}^{(q)}(11';\beta, \mu)&=& \frac{1}{i}\langle \Psi(1)
\Psi^{\dagger}(1') \rangle_{q} \;,\nonumber \\
\tilde{G}_{<}^{(q)}(11';\beta, \mu)&=&\frac{\pm}{i} \langle
\Psi^{\dagger}(1')\Psi(1) \rangle_{q} \;. \label{9}
\end{eqnarray}
The notation $>$ and $<$ is intended to exhibit the feature that
$\tilde{G}^{(q)}(1,1';\beta, \mu)=\tilde{G}_{>}^{(q)}(1,1';\beta,
\mu)$ for $t_{1}\! >\! t_{1'}$ and $\tilde{G}^{(q)}(1,1';\beta,
\mu)=\tilde{G}_{<}^{(q)} (1,1';\beta, \mu)$ for $t_{1} < t_{1'}$.
Using (\ref{7qq}), we may similarly express $\tilde{G}_{>}^{(q)} $
and $\tilde{G}_{<}^{(q)}$ in terms of the corresponding grand
canonical correlation functions. The spectral weight function in
frequency space by taking the Fourier transform with respect to
time differences, $A(\vec{r}_{1}, \vec{r}_{1'};\omega)$,
introduced in KB reflects only the properties of the Hamiltonian
$\hat{H}$. The average occupation number in the grand canonical
ensemble of a mode with energy $\omega$, $f(\omega,\beta)=
(\exp(\beta (\omega -\mu)) \mp 1)^{-1}$, takes account of the
basic permutation symmetry of the system. We can thus express
$\tilde{G}_{>}^{(q)} $ and $\tilde{G}_{<}^{(q)}$ in terms of these
in the following way:
\begin{eqnarray}
& &i \tilde{G}_{>}^{(q)}(\vec{r}_{1}, \vec{r}_{1'};\omega;\beta,
\mu)= \int_{C} {\mbox {d}}u \tilde{K}_q^{(2)}(u)
\tilde{Z}_{1}(-u(1-q) \tilde{\beta}, \mu) \;i
\tilde{G}^{(1)}_{>}(\vec{r}_1,\vec{r}_{1'};\omega ;
-u(1-q)\tilde{\beta}, \mu)\; \nonumber \\&=&\int_{C}  {\mbox {d}}u
\tilde{K}_q^{(2)}(u)(1 \pm f(\omega,-u(1-q)\tilde{\beta}, \mu))
A(\vec{r}_{1}, \vec{r}_{1'}; \omega)
\tilde{Z}_{1}(-u(1-q)\tilde{\beta},\mu) \;\; , \label{10}
\end{eqnarray}
\begin{eqnarray}
& &i \tilde{G}_{<}^{(q)}(\vec{r}_{1},
\vec{r}_{1'};\omega;\beta,\mu)= \int_{C} {\mbox {d}}u
\tilde{K}_q^{(2)} (u)\tilde{Z}_{1}(-u(1-q)\tilde{\beta}, \mu)\;i
G^{(1)}_{<}(\vec{r}_1,\vec{r}_{1'};\omega ;-u(1-q)\tilde{\beta}
,\mu)\; \nonumber
\\&=&\pm \int_{C}  {\mbox {d}} u \tilde{K}_q^{(2)}(u)
f(\omega,-u(1-q)\tilde{\beta}, \mu)A(\vec{r}_{1},
\vec{r}_{1'};\omega) \tilde{Z}_{1}(-u(1-q)\tilde{\beta},\mu) \;\;.
\label{11}
\end{eqnarray}
Thus the spectral weight function is found to be
\begin{eqnarray}
A(\vec{r}_1, \vec{r}_{1'} ; \omega) &= &i \left(
\tilde{G}_{>}^{(q)} (\vec{r}_{1}, \vec{r}_{1'};\omega;\beta, \mu)
- \tilde{G}_{<}^{(q)}(\vec{r}_{1}, \vec{r}_{1'}; \omega;\beta,
\mu) \right)\nonumber \\ &=& \int_{C}  {\mbox du}
\tilde{K}_q^{(2)} (u)A(\vec{r}_{1}, \vec{r}_{1'};\omega)
\tilde{Z}_{1}(-u(1-q)\tilde{\beta}, \mu) \;\;. \label{12}
\end{eqnarray}
From this we deduce an important sum rule
\begin{eqnarray}
& &i  \int_{-\infty}^{\infty} \frac{{\mbox d\omega}}{2 \pi} \left(
\tilde{G}_{>}^{(q)}(\vec{r}_{1}, \vec{r}_{1'};\omega;\beta, \mu) -
\tilde{G}_{<}^{(q)}(\vec{r}_{1}, \vec{r}_{1'};\omega;\beta, \mu )
\right)= \delta \left( \vec{r}_{1}- \vec{r}_{1'} \right) \;\;.
\label{13}
\end{eqnarray}
This is just an expression of the equal time CCR of the particle
fields. The spectral function weight is a property of the given
systems {\it not dependent on the ensemble} and the results have
similar appearance as in \cite{RAJA2} with the modifications noted
here. Using the Fourier representations of the step functions
involved in the time-ordered Green function in Eq.(\ref{6qq}), an
important result, generalizing the result found in \cite{Abe2} is
deduced:
\begin{eqnarray}
& &\tilde{G}^{(q)}(1,1' ;\beta, \mu )= \int_{C} {\mbox du}\;
\tilde{K}_q^{(2)}(u) \tilde{Z}_{1}(-u(1-q)\tilde{\beta}, \mu )\;
\int_{-\infty}^{\infty}\frac{{\mbox {d}} \omega}{2 \pi} e^{-i
\omega(t_1 -t_{1'})} \times \nonumber \\ &\times&\!\!
\int_{-\infty}^{\infty} \frac{{\mbox {d}} \omega'}{2 \pi}
A(\vec{r}_1,\vec{r}_{1'};\omega') \left \{
P\left(\frac{1}{\omega-\omega'} \right)\! -\! i\pi \delta(\omega
-\omega') \left[ \tanh \left(\frac{-u}{2}\tilde{\beta}(1-q)
 (\omega -\mu) \right)\right]^{\mp 1} \right \}\;.
\label{8.4qq}
\end{eqnarray}
Note that in the view of Eq.(\ref{12}), only the delta function
term in this equation depends explicity of $q$ as in
Ref.\cite{Abe2}. So far we have discussed the one-particle
properties. The above development is similarly extended to
generalize the many-particle $q$-Green functions. Using the same
notations as in KB, we have in general,
\begin{eqnarray}
& &\tilde{G}_{n}^{(q)}(12...n, 1'2'...n'; \beta, \mu) =
\frac{1}{i^n}\langle T (
\Psi(1)\Psi(2)...\Psi(n)\Psi^{\dagger}(1')
\Psi^{\dagger}(2')...\Psi^{\dagger}(n') )\rangle_q \nonumber \\
&=& \int_{C} {\mbox du} \tilde{K}_q^{(2)} (u)
\tilde{Z}_{1}(-u(1-q)\tilde{\beta}, \mu ) \;G_n^{(1)}(12...n,
1'2'...n'; -u(1-q)\tilde{\beta}, \mu) \;. \label{1111111116}
\end{eqnarray}
For a many-particle system with a Hamiltonian containing one-body
potential $V_1(\vec{r}_1)$ and instantaneous two-body interaction
potential $V_2(\vec{r}_1, \vec{r}_2)$, which is symmetric under
interchange of $1 $ and $2$:
\begin{eqnarray}
\hat{H} &=& \int {\mbox {d$\vec{r}$}}\; \frac{\nabla
\Psi^{\dagger}(\vec{r},t)\cdot \nabla \Psi(\vec{r},t)}{2m}
+\int{\mbox {d$\vec{r}$}}\; V_1(\vec{r})\Psi^{\dagger}(\vec{r},t)
\Psi(\vec{r},t))\nonumber \\ &+& \frac{1}{2}\int\int{\mbox
{d$\vec{r}$}}{\mbox {d$\vec{r}\;'$}}
\;\Psi^{\dagger}(\vec{r},t)\Psi^{\dagger}(\vec{r}\;',t)
V_2(\vec{r}, \vec{r}\;')\Psi(\vec{r}\;',t)\Psi(\vec{r},t)\;.
\label {H11}
\end{eqnarray}
We can express $\tilde{U}_q$ in terms of the Green function
following KB
\begin{eqnarray}
\tilde{U}_q = \frac{ \pm i}{4} \!\int \!\! {\mbox {d$\vec{r}$}} \;
\left[i \left( \frac{\partial } {\partial t} - \frac{\partial
}{\partial t'} \right) +  \frac{\nabla \cdot \nabla'}{m} -V_{1}(r)
- V_{1}(r')\right]\tilde{G}_{<}^{(q)}(\vec{r},t;\vec{r}\;',t')
\Big|_{\vec{r}\;'=\vec{r}, t'=t } \;.
\end{eqnarray}
For the free particle case $ V_{1}(r)$ $=$ $0 $ as in
Eqs.(\ref{14}), (\ref{15}) we have,
\begin{eqnarray}
\frac{\tilde{U}_q}{V} &=& \int_{C}{\mbox du}
\tilde{K}_q^{(2)}(u)\int_{-\infty}^{\infty} \frac{ {\mbox d
\omega}}{2 \pi} \int \frac{{\mbox d^{D} p}} {(2 \pi )^{D}}
\left(\frac{ \omega+ p^2/2m}{2} \right) \frac{\tilde{Z}_{1}
(-u(1-q)\tilde{\beta}, \mu )}{(e^{-u(1-q)\tilde{\beta} ( \omega
-\mu)} \mp 1)} A( \vec{p} ; \omega) \;. \label{111111116}
\end{eqnarray}
For a uniform system, we can take Fourier transforns  with respect
to $\vec{r}_{1} -\vec{r}_{1'}$ in Eq. (\ref{11}) and express the
one-particle momentum distribution function $\langle
\hat{N}(\vec{p}\;) \rangle_q$ in terms of the spectral  weight
function of the N-particle system.
\begin{eqnarray}
\tilde{N}(\vec{p} \;)_q = \pm \int_{C}{\mbox {d}}u
\tilde{K}_q^{(2)}(u)\int_{-\infty}^{\infty} \frac{{\mbox d
\omega}}{2 \pi} \frac{A(\vec{p}; \omega)
\tilde{Z}_{1}(-u(1-q)\tilde{\beta} ,\mu)}
{(e^{-u(1-q)\tilde{\beta}( \omega -\mu)} \mp 1)} \;. \label{14}
\end{eqnarray}
Similarly the one-particle frequency distribution function
$\langle \hat{N}(\omega )\rangle_q$ is given by
\begin{eqnarray}
\tilde{N}( \omega )_q &=& \pm V \int_{C}{\mbox du}
\tilde{K}_q^{(2)}(u) \frac{\tilde{Z}_{1}(-u(1-q)\tilde{\beta},
\mu)} {(e^{-u(1-q)\tilde{\beta}( \omega -\mu)} \mp 1)} \int
\frac{{\mbox d^{D}p}}{(2 \pi )^D} A(\vec{p} ; \omega) \;.
\label{15}
\end{eqnarray}
Here $V$ is the  volume of the $D$ dimensional space in which the
particles reside. The chemical potential is determined by the
expression for the $q$-mean value of the total number operator
$\hat{N}$ ,
\begin{eqnarray}
\frac{\tilde{N}_q}{V} = \pm \int_{C}{\mbox du}
\tilde{K}_q^{(2)}(u) \int_{-\infty}^{\infty} \frac{ {\mbox d
\omega}}{2 \pi} \int \frac{{\mbox d^{D} p}}{(2 \pi)^{D}}
\frac{\tilde{Z}_{1} (-u(1-q)\tilde{\beta}, \mu
)}{(e^{-u(1-q)\tilde{\beta} ( \omega -\mu)} \mp 1)} A( \vec{p} ;
\omega) \;. \label{16}
\end{eqnarray}

Equation (\ref{14}) will now be used to derive the one-particle
$q$-momentum distribution function for a free fermi gas, where
$A(\vec{p}, \omega)= 2 \pi \delta \left(\omega - \epsilon_p
\right)$, $\epsilon_p$ being the one-particle energy. The known
partition function for the free fermi gas in the Gibbs theory need
in this expression is approximated in a simple form
\begin{eqnarray}
\ln Z_1(\beta, \mu) = V \left( \frac{m}{2 \pi \beta} \right)^{3/2}
\sum^{\infty}_{l=1} \frac{(-1)^{(l+1)}e^{\beta \mu l}}{l^{5/2}}
\cong V \left( \frac{m}{2 \pi \beta} \right)^{3/2}e^{\beta \mu} \;
,
\end{eqnarray}
and hence also a further approximation resembling the fugacity
expansion of the classical Maxwell gas,
\begin{eqnarray}
Z_1(\beta, \mu) \cong  \sum^{\infty}_{l=0} \frac{1}{l!} \left[ V
\left( \frac{m}{2 \pi \beta} \right)^{3/2} \right]^l e^{\beta \mu
l} \;,
\end{eqnarray}
and $1/(e^x +1)=\sum_{J=1}^{\infty}(-1)^{J+1}e^{-Jx}$ for
$|e^{-x}|<1$, we have finally the result,
\begin{eqnarray}
\frac{\tilde{N}(\vec{p} \;)_q}{V} &\cong& \frac{\Gamma\left(
\frac{1}{1-q} \right)}{\tilde{Z}_q}\sum_{l=0}^{\infty}\frac{1}{l!}
\left[V \left( \frac{m}{2 \pi \tilde{\beta}} \right)^{3/2}
\right]^{l} \nonumber \\ &\times&
\sum_{J=1}^{\infty}\frac{(-1)^{J+1}}{\Gamma\left( \frac{1}{1-q}+
\frac{3l}{2}\right)} \left \{  1+(1-q)\tilde{\beta}\left(
\tilde{U}_q -\mu\tilde{N}_q +\mu l -J\left(\epsilon_p -\mu\right)
\right)\right\}^{q/(1-q)+3l/2} \;.
\end{eqnarray}
An approximate expression for the $q$-mean momentum distribution
function in the first version of the Tsallis formulation
\cite{NORMA} (where ordinary $q$-mean values were used) has been
used in the literature \cite{T1}. We can deduce such an expression
as in \cite{T1} from the above by making a further drastic
approximation by taking $l=0$,$\tilde{Z_q}=1$, dropping
$\tilde{U}_q$, $\tilde{N}_q$ and $\tilde{\beta}=\beta$ to be in
conformity with the first version of the Tsallis formalism
\cite{NORMA}, we obtain the result,
\begin{equation}
\frac{\tilde{N}(\vec{p} \;)_q}{V} \approx \sum_{J=1}^{\infty}
(-1)^{J+1} \left \{1-(1-q)\beta J\left(\epsilon_p -\mu \right)
\right\}^{q/(1-q)} \;. \label{E1}
\end{equation}
A futher approximation corresponding to the high temperature
scheme where $\beta \left( \epsilon_p -\mu \right)<<1$, we have
\begin{equation}
\frac{\tilde{N}(\vec{p} \;)_q}{V} \approx \sum_{J=1}^{\infty}
(-1)^{J+1}\left \{1-(1-q)\beta \left(\epsilon_p -\mu \right)
\right\}^{qJ/(1-q)}= \frac{1}{1+ \left \{ 1-(1-q)\beta
\left(\epsilon_p -\mu \right) \right\}^{-q/(1-q)}} \;. \label{E2}
\end{equation}
In order to assess the accuracy of the approximation in
Eq.(\ref{E2}) in relation to the more exact result in
Eq.(\ref{E1}), we compare the respective second order terms of the
series, $(qJ(qJ-(1-q))/2)\left(\beta \left( \epsilon_p -\mu
\right)\right)^2$ and $(q(2q-1)/2)\left(J\beta \left( \epsilon_p
-\mu \right)\right)^2$, and we find the error involved is of order
$(1-q)$. This result was essentially obtained in \cite{T1} via a
different approach to this problem. Similar calculations for the
Bose and the Maxwell gases may be made on the same lines as above
yielding corresponding expressions for the respective momentum
distribution functions.

${ }$

{\section {Linear Response Functions}}

The linear response is now calculated, following \cite{R96}, using
this new prescription:
\begin{eqnarray}
\langle \Delta \hat{B}(t) \rangle_q \equiv {\mbox {Tr}} \left[
\hat{P}(t) \hat{B} \right] - {\mbox {Tr}} \left[
\hat{\tilde{P}}(\hat{H};q,\tilde{\beta}) \hat{B}\right]
\label{k11}
\end{eqnarray}
where $\hat{P}(t)=\hat{\rho}^q(t)/{\mbox {Tr}} \hat{\rho}^q(t)$.
With this new definitions, all averages are calculated with proper
unit normalization. The time-dependent $P$-operator in
Eq.(\ref{k11}) obeys the usual equation of motion with
time-dependent Hamiltonian $\hat{H} -\hat{A} X(t)$
\begin{eqnarray}
i \hbar \frac{\partial }{\partial t} \hat{P}(t)= \left[\hat{H},
\hat{P}(t) \right] - \left[\hat{A}, \hat{P}(t)\right]X(t)
\label{k12}
\end{eqnarray}
with the initial condition $\hat{P}(t= -\infty) =
\hat{\tilde{P}}(\hat{H};q,\beta)$ given by Eq.(\ref{1qq}). Taking
the trace over both sides, we observe that ${\mbox {Tr}}
\hat{P}(t)=1$ for all times. The solution to this linear order in
$X(t)$ is the found the standard procedures:
\begin{eqnarray}
\hat{P}(t) \approx \hat{\tilde{P}}(\hat{H};q,\beta) -\frac{1}{i
\hbar} \int_{-\infty}^{t} {\mbox {d}}t e^{-i(t-t')\hat{H}/\hbar}
\left[\hat{A},\hat{\tilde{P}}( \hat{H};q,\tilde{\beta}) \right]
e^{i(t-t')\hat{H}/\hbar} \;.
\end{eqnarray}
Thus
\begin{eqnarray}
\langle \Delta \hat{\tilde{B}} \rangle_q&=& \int_{-\infty}^{t}
{\mbox {d}} t  \tilde{\phi}_{BA}^{(q)}(t-t')X(t') \nonumber \\
\tilde{\phi}_{BA}^{(q)}(t-t')&=& -\frac{1}{i \hbar}{\mbox {Tr}}
\left\{ \left[\hat{A},
\hat{\tilde{P}}(\hat{H};q,\beta)\right]\hat{B}(t) \right\}
\nonumber \\ &=& \frac{1}{i \hbar} {\mbox {Tr}} \left\{ \left[
\hat{A},\hat{B}(t) \right] \hat{\tilde{P}} (\hat{H};q,\beta)
\right\} \; .
\end{eqnarray}
It appears therefore all the results and conclusions of \cite{R96}
hold with the new definition of mean values with the
renormalization of the temperature and the self-consistency
condition in Eq.(\ref{1qq},\ref{2qq}).

We now turn our attention to rewriting the dynamic response and
the scattering cross section in the $q$-formalism in terms of the
integrals over the usual ones as was done above. Reformulating the
result obtained in Ref.\cite{R96}, the dynamic linear response of
a quantity $\hat{B}$ to an external probe that generates $\hat{A}$
in the $q$-formalism is
\begin{eqnarray}
\tilde{\chi}_{BA}^{(q)}(\omega, \beta, \mu)= \lim_{\epsilon
\rightarrow 0} \int_{0}^{\infty} {\mbox {dt}} \; e^{-i \omega t
-\epsilon t} \frac{1}{i}\tilde{\phi}_{BA}^{(q)}(t,\beta, \mu)
\end{eqnarray}
This in terms of the integral representation, is:
\begin{eqnarray}
\tilde{\chi}_{BA}^{(q)}(\omega, \beta, \mu)&=& \int_{C}{\mbox
{d}}u \tilde{K}_q^{(2)} (u) \tilde{Z}_{1}(-u(1-q)\tilde{\beta},
\mu) \chi _{BA}^{(1)}(\omega,-u(1-q)\tilde{\beta}, \mu)
\end{eqnarray}
where $\chi _{BA}^{(1)}(\omega,-u(1-q)\tilde{\beta}, \mu)$ is the
usual Kubo dynamical response function evaluated now at a
temperature $-u(1-q)\tilde{\beta}$. Following Ref.\cite{R96} in
the context above and rewriting the $q$-averages  of the
anticommutator and commutator expressions, we have:
\begin{eqnarray}
\tilde{\Psi}_{BA}^{(q)}(t, \beta, \mu)&=& \frac{1}{2} {\mbox {Tr}}
\left[ \hat{\tilde{P}}(\hat{H},\hat{N};q,\beta, \mu) \left[
\hat{A}(0) \hat{B}(t)   +  \hat{B}(t) \hat{A}(0) \right] \right]
\nonumber \\ & = & \frac{1}{2} \int_{C}{\mbox
du}\tilde{K}_{q}^{(2)}(u) Z_{1}(-u(1-q)\tilde{\beta},
\mu)\Psi_{BA}^{(1)}(t, -u(1-q) \tilde{\beta}, \mu)
\end{eqnarray}
\begin{eqnarray}
\tilde{\Phi}_{BA}^{(q)}(t, \beta) &=& \lim_{\epsilon \rightarrow
0} \int^{\infty}_{t}\!\!\!\!\! {\mbox dt'} e^{-\epsilon t'} {\mbox
{Tr}} \left[\hat{\tilde{P}} ( \hat{H},\hat{N};q,\beta, \mu)
\left[\hat{A}(0),\hat{B}(t) \right] \right] \nonumber \\ &=&
\int_{C} {\mbox {d}}u
\tilde{K}_{q}^{(2)}(u)\tilde{Z}_{1}(-u(1-q)\tilde{\beta}, \mu)
\Phi _{BA}^{(q=1)}(t',-u(1-q)\tilde{\beta}, \mu)  \; .
\end{eqnarray}
The  fluctuation-dissipation theorem due to Kubo \cite{Kubo} for
the extensive case, $(q=1)$ is
\begin{eqnarray}
\Psi^{(1)}_{BA}(\omega,\beta, \mu )=E_{\beta}(\omega) \Phi
_{BA}^{(1)}(\omega,\beta, \mu)
\end{eqnarray}
with
\begin{eqnarray}
E_{\beta}(\omega)=\frac{\omega}{2} \coth\left(\frac{\beta
\omega}{2} \right)\;.
\end{eqnarray}
Here we obtain
\begin{eqnarray}
\tilde{\Psi}_{BA}^{(q)}(\omega, \beta, \mu) &=& \frac{\omega}{4}
\int_{C}{\mbox du} \tilde{K}_{q}^{(2)}(u)
\tilde{Z}_{1}(-u(1-q)\tilde{\beta}, \mu) \nonumber \\ &\times&
\coth\left(-u(1-q)\tilde{\beta} \right)\Phi_{BA}^{(1)}(\omega,
-u(1-q)\tilde{\beta}, \mu)) \;.
\end{eqnarray}
We now relate the scattering function defined for example in,
Lovesey\cite{SWL}, in the $q$-formalism as
\begin{eqnarray}
\tilde{S}^{(q)} (\vec{k}, \omega, \beta) = \frac{1}{2 \pi} \int
_{-\infty}^{\infty} {\mbox {dt}} \exp(-i \omega t) \langle
\hat{A}^{\dagger}(0) \hat{A}(t)\rangle_{q}^{(c)} \;,
\label{24242424}
\end{eqnarray}
where $ \hat{A} $ is the operator which affects the change in the
states of the system in a scattering process. Here the superscript
$(c)$ denotes canonical ensemble instead of the grand canonical
ensemble used earlier. This is equivalent formally to setting $\mu
= 0$ in the earlier development. Then, using our transformation,
we express this scattering function in terms of the usual $q=1$
scattering function
\begin{eqnarray}
\tilde{S}^{(q)} (\vec{k}, \omega, \beta) &=& \int _{C} {\mbox
{d}}u \tilde{K}_{q}^{(2)}(u) \tilde{Z}_1(-u(1-q)\tilde{\beta})
S^{(1)}(\vec{k}, \omega , -u(1-q)\tilde{\beta})\;. \label{25}
\end{eqnarray}
From \cite{R96}, by taking $\hat{B}=\hat{A}^{\dagger}$, we have
that the imaginary part of the $q$-susceptibility,
$\tilde{\chi}^{(q)}_{\hat{A}^{\dagger} \hat{A}}( \omega, \beta)$
can be expressed in terms of the $q=1$ scattering function
\begin{eqnarray}
& &{\mbox {Im}}\;\tilde{\chi}^{(q)}_{\hat{A}^{\dagger}\hat{A}} (
\vec{k}, \omega, \beta) = \pi \int _{C} {\mbox {du}}
\tilde{K}_{q}^{(2)}(u)\tilde{Z}_1 (-u(1-q)\tilde{\beta})\times
\nonumber \\ &\times& (1- exp(-u(1-q)\tilde{\beta})
S^{(1)}(\vec{k},\omega , -u(1-q)\tilde{\beta}) \;.
\end{eqnarray}
We have thus expressed the $q$-scattering function as well as the
imaginary part of the associated $q$-susceptibility in terms of
the parametric integrals over a kernel multiplied by the usual
scattering function which now depends on this parameter as
displayed above. It is possible to obtain other properties as in
Ref. \cite{R96} to this new version of the Tsallis statistics.

${ }$ {\section {BOLTZMANN TRANSPORT-LIKE EQUATION}}

There exists a class of disturbances which are not conveniently
described by the usual equilibrium Green's function \cite{K62}.
For example, the disturbance produced by the externally applied
force field, $\hat{F}(\vec{r},t)=\nabla U(\vec{r},t)$. And many
interesting physical phenomena appear as response of systems to
external disturbances of this kind, for example in an ordinary
gas, a slowly varying $U(\vec{r},t)$ produces sound waves. Thus,
we expect that this feature will appear in the generalized
$q$-ensemble theory as in $q=1$. It is interesting therefore to
extend the usual Boltzmann equation to the generalized case to
understand this in the nonextensive context. In our formalism,
this disturbance (force) may be represented by,
\begin{eqnarray}\hat{H}'(t)= \int {\mbox {d}}^3r\;
\hat{n}(r,t)U(r,t)
\end{eqnarray}
where $\hat{n}(r,t)=\Psi_{U}^{\dagger}(r,t)\Psi_{U}(r,t)$. Let us
to obtain the colisionless Boltzmann equation from the Green
functions given above using the Hartree approximation in the same
spirit as KB. By using the Heisenberg notation we have that the
$q$-expectation average is given by
\begin{eqnarray}
\langle  \hat{X}({\bf R},t) \rangle_{U,q}=
\frac{\sum_i\hat{\tilde{\rho}}_{i}^q \langle i,t_0|\hat{X}_U({\bf
R},t) |i,t_0 \rangle}{\sum_i \hat{\tilde{\rho}}_{i}^q}
\end{eqnarray}
where
\begin{eqnarray}\hat{X}_U({\bf R},t)={\cal V}(t)^{-1}\hat{X}({\bf R},t){\cal V}(t)
\end{eqnarray}
with
\begin{eqnarray}
{\cal V}(t)={\mbox {T}}\left \{ \exp \left [ -i \int_{t_0}^{t}
{\mbox {d}}t'\int {\mbox {d}}^3r' \hat{n}(r',t')U(r',t') \right]
\right \} \;\;.
\end{eqnarray}
From the Eq.(\ref{9}) we have that,
\begin{eqnarray}\tilde{G}_{<}^{(q)}(1,1';U)= \pm \frac{1}{i}
\langle {\mbox {T}}(\Psi_U^{\dagger}(1')\Psi_U(1)) \rangle_q
\end{eqnarray}
\begin{eqnarray}
\tilde{G}^{(q)}(12,1'2';U)= \left(\frac{1}{i} \right)^2
\langle{\mbox {T}}
(\Psi_U(1)\Psi_U(2)\Psi_U^{\dagger}(2)\Psi^{\dagger}_U(1))
\rangle_q \;\; .
\end{eqnarray}
In terms of these Green's functions we may describe the response
of a system, initially in thermodynamic equilibrium, to an applied
disturbance $U(\vec{r},t)$. In the same way as above, the average
density at point ${\bf R},T$ is
\begin{eqnarray}
\langle \hat{n}({\bf R},T) \rangle_{U,q} &=&\langle
\Psi^{\dagger}_U({\bf R},T) \Psi_U({\bf R},T) \rangle_q\nonumber
\\ &=& \pm i\tilde{G}_{<}^{(q)} ({\bf R},T,{\bf R},T;U)
\end{eqnarray}
and the current at the same point is
\begin{eqnarray}
\langle \hat{J}({\bf R},T) \rangle_{U,q} = \left \{ \frac{\nabla
-\nabla'}{2mi} \left[ \pm i \tilde{G}_{<}^{(q)}({\bf R},T,{\bf
R}',T;U) \right] \right \}_{{\bf R}={\bf R}'} \;\;,
\end{eqnarray}
where the conservation laws for the number of particles, the
energy and the momentum are preserved here as well as in the usual
case, and we can use them for derivation of the sound propagation.
Here we define $f_q({\bf p}, {\bf R}, T)$ (with
${\bf{r}}=\vec{r}_1-\vec{r}_{1'}$, ${\bf
R}=1/2(\vec{r}_1+\vec{r}_{1'})$ $t=t_1-t_{1'}$ and
$T=1/2(t_1+t_{1'})$) as,
\begin{eqnarray}
f_q({\bf p}, {\bf R}, T)&=& \int \frac{{\mbox {d}}\omega}{2\pi}
\tilde{G}_{<}^{(q)}({\bf p},\omega, {\bf R},T;U)\nonumber \\ &=&
\int {\mbox {d}}^3 { r}e^{-i{\bf p}\cdot {\bf r}} \langle
\Psi_U^{\dagger}\left( {\bf R}-\frac{{\bf r}}{2}, T \right) \Psi_U
\left({\bf R}+\frac{{\bf r}}{2}, T \right) \rangle_q
\label{final3}
\end{eqnarray}
generalizing the original definition proposed by Wigner. As in the
usual case $f_q({\bf p}, {\bf R}, T)$ leads to the generalized
$q$-particle density
\begin{eqnarray}
\int \frac{{\mbox {d}}^3p}{(2 \pi)^3}f_q({\bf p}, {\bf R}, T)=
\langle \Psi_U^{\dagger}\left( {\bf R}, T \right)\Psi_U \left({\bf
R},T \right) \rangle_q=\langle \hat{n}({\bf R},T) \rangle_q
\end{eqnarray}
and the generalized $q$-particle current
\begin{eqnarray}
\langle { \hat{J}({\bf R},T)} \rangle_{U,q}= \int \frac{{\mbox
{d}}^3 p}{(2 \pi)^3}\frac{{\bf p}}{m} f_q({\bf p}, {\bf R}, T)
\;\; .
\end{eqnarray}
This identification above of the distribution function $f_q({\bf
p}, {\bf R}, T)$ will enable us to see the relationship between to
Green's functions, transport equations, and the colisionless
Boltzmann equation. Thus, to obtain the colisionless Boltzmann
equation we use
\begin{eqnarray}
& &\left( i\frac{\partial }{\partial t_1}+\frac{\nabla^2_1} {2m}
-U(1)\right)\tilde{G}^{(q)}(1,1';U)= \delta(1-1') \nonumber \\
&\pm& \int\int {\mbox {d}}t_2 {\mbox
{d}}r_2v({\vec{r}}_1-{\vec{r}}_2)
\delta(t_1-t_2)\tilde{G}^{(q)}(12,1'2^{+};U)
\end{eqnarray}
and
\begin{eqnarray}
\tilde{G}^{(q)}(12,1'2';U)=\tilde{G}^{(q)}(1,1';U)\tilde{G}^{(q)}(2,2';U)
\;\;.
\end{eqnarray}
After some simplification, we obtain
\begin{eqnarray}\left( i\frac{\partial }{\partial t_1}+\frac{\nabla^2_1}{2m} -
U_{eff}(1)\right)\tilde{G}^{(q)}(1,1';U)= \delta(1-1')
\end{eqnarray}
where
\begin{eqnarray}
U_{eff}({\bf R},T)=U({\bf R},T) \pm i \int {\mbox {d}}{\bf R'}
v({\bf R}-{\bf R'})\tilde{G}_{<}^{(q)}({\bf R'},T;{\bf R'},T) \;\;
.
\end{eqnarray}
By taking the difference of two equations in the variables $1$ and
$1'$, we find
\begin{eqnarray}
\left \{ i \!\left(\frac{\partial }{\partial
t_1}\!+\!\frac{\partial } {\partial t_{1'}}
\right)+(\nabla_1+\nabla_{1'})\cdot \left(
\frac{\nabla_1-\nabla_{1'}}{2m} \right)-\left[ U_{eff}(1)-
U_{eff}(1') \right] \right \} \tilde{G}^{(q)}(1,1';U)= 0 \;\; .
\label{final1}
\end{eqnarray}
Considering now $t_{1'}=t^{+}_{1}=T$ and expressing
Eq.(\ref{final1}) in terms of, ${\bf r}=\vec{r}_1-\vec{r}_{1'}$,
${\bf R}=1/2(\vec{r}_1+\vec{r}_{1'})$ and using Eq.(\ref{final3})
we have that
\begin{eqnarray}
&\pm&\left \{ \frac{\partial } {\partial T}+\frac{\nabla_{{\bf
R}}\cdot \nabla_{{\bf r}}}{im} -\frac{1}{i} \left[U_{eff}({\bf
R}+\frac{{\bf r}}{2},T)- U_{eff}({\bf R}-\frac{{\bf r}}{2},T)
\right] \right \} \nonumber \\ &\times& \int\frac{{\mbox {d}}^3
p'}{(2 \pi)^3}e^{i{\bf p}'\cdot{\bf r}}f_q({\bf p'}, {\bf R}, T)=
0 \;\; . \label{final}
\end{eqnarray}
Now let us suppose that $U_{eff}({\bf R},T)$ varies slowly in
${\bf R}$, then we obtain after some calculation the collisionless
Bolztmann equation
\begin{eqnarray}
\left[ \frac{\partial }{\partial T}+\frac{{\bf p}}{2m}\cdot
\nabla_{{\bf R}} -\nabla_{{\bf R}}U_{eff}({\bf
R},T)\cdot\nabla_{{\bf p}} \right] f_q({\bf p}, {\bf R}, T)= 0
\end{eqnarray}
where
\begin{eqnarray}
U_{eff}({\bf R},T)=U({\bf R},T)+ \int {\mbox {d}}{\bf R}' v({\bf
R}-{\bf R}')\int \frac{{\mbox {d}}^3p'} {(2 \pi)^3}f_q({\bf
p'},{\bf R}',T) \;.
\end{eqnarray}

${ }$ {\section {Summary and Conclusions}} In this paper we have
developed in detail the Green function theory for nonextensive
systems based on the $q$-ensemble of Tsallis. By means of a
contour representation, Eq.(\ref{7qq}), we have made this theory
resemble the usual one for extensive systems given by L. P.
Kadanoff and G. Baym\cite{K62}, for example, even though in actual
practice, the results are very different, as exemplified by the
representative results given in Sec. II for a variety of
situations. Before this development, thermodynamic quantities for
model systems were computed in the Tsallis ensemble as for example
in Ref.\cite{A1,17.5}. With the present work, we believe that the
theory of many particle systems for the Tsallis ensemble has been
considerably extended and placed on par with the conventional
theory based on the Gibbsian ensemble, in that we have been able
to compute response functions besides the thermodynamic
quantities. The case of $q$ different from unity is expected to
apply for long-range interacting Hamiltonian systems\cite{NORMA},
among others. From a formal point of view, noninteracting and
short-range interacting systems are mathematically well posed
problems only for $q \leq 1$. In conclusion, we have here
developed the formalism associated with Tsallis statistics for
describing nonextensive many-particle systems by a suitable
generalization of the corresponding Green function techniques, so
commonly employed in such studies for extensive systems. As with
the usual Green function theory, which has been traditionally
successful in explaining experimental observations, the present
work enables us to analyze future possible experimental work on
nonextensive systems. We may add that other forms of the
nonextensive entropy such as $S^{(a)}_q= S_q/c_q$, deduced from
considerations of form invariance \cite{R2} of the statement of
maximum entropy principle and the metric structure in quantum
density matrix theory, many also be expressed in contour integral
form. The work presented here thus admits of applications
elsewhere.

{\section* {Acknowledgements}} AKR thanks the Office of Naval
Research for partial support of this work. EKL acknowledges
partial support from CNPq and PRONEX (Brazilian agencies).


\references
\bibitem{RAJA}
A. K. Rajagopal, R. S. Mendes and E. K. Lenzi, Phys. Rev.Lett.
{\bf 80}, 3907 (1998)
\bibitem{RAJA2}
E. K. Lenzi, R. S. Mendes and A. K. Rajagopal ,Phys. Rev. E {\bf
59}, 1397 (1999).
\bibitem{NORMA}
C. Tsallis, R. S. Mendes and A. R. Plastino, Physica A {\bf 261},
534 (1998).
\bibitem {T88}
C. Tsallis, {J. Stat. Phys.} {\bf 52}, 479 (1988); see also Chaos,
Solitons and Fractals, {\bf 6}, 539 (1995).
\bibitem {CT91}
E. M. F. Curado  and  C. Tsallis,  {J. Phys. A} {\bf 24}, L69
(1991); Errata: {\bf 24}, 3187 (1991); {\bf 25}, 1019 (1992); See
http://tsallis.cat.cbpf.br/biblio.htm for a periodically updated
bibliography on the subject.
\bibitem {R96}
A. K. Rajagopal, {Phys. Rev. Lett.} {\bf 76}, 3469 (1996).
\bibitem{Abe1}
S. Abe, {\it Thermodynamics limit and classical ideal gas in
nonextensive statistical mechanics with normalized $q$-expectation
values}, preprint (1998).
\bibitem{Abe2}
S. Abe, {\it Thermal Green functions in nonextensive statistical
mechanics }, Eur. Phys. J. B. (1999) in press.
\bibitem {A1}
A. R. Plastino and A. Plastino and C. Tsallis, J. Phys. A {\bf
27}, 5707 (1994).
\bibitem {K62}
L. P. Kadanoff and G. Baym, {\it Quantum Statistical Mechanics},
(W. A. Benjamin, Inc., New York, 1962).
\bibitem{T1}
F. B\"uy\"ukkil\c{c}, D. Demirhan, and A. G\"ule{c}, Phys. Lett. A
{\bf 197}, 209 (1995).
\bibitem{Kubo}
R. Kubo, J. Phys. Soc. Jpn. {\bf 12}, 570 (1957).
\bibitem{SWL}
S. W. Lovesey, {\it  Condensaded Matter  Physics}, (The
Benjamin/Cummings Plublishing Company, Massachusetts, 1980).
\bibitem{17.5}
S. Curilef, Phys. Lett.  A {\bf 218}, 11 (1996); see also S.
Curilef, Z. Phys. B {\bf 100}, 433 (1996).
\bibitem{R2}
A. K. Rajagopal and Sumiyoshi Abe, submitted for publication
(1999).
\end{document}